\documentclass[12pt,times,endnote]{article}
\usepackage{epsfig,float,color,graphicx}
\usepackage{amsmath}
\usepackage{amssymb}
\newcommand{\re}{\mathbb R}
\usepackage{amscd}
\begin{document}
\begin{center}
{\large\bf Brownian motion and magnetism}\\*[0.8cm]
Supurna Sinha\\*[0.2cm]
{\it Centre for Theoretical Studies, Indian Institute of Science,\\ 
Bangalore, India 560 012}\\*[0.6cm]
Joseph Samuel \\*[0.2cm]
{\it Raman Research Institute, Bangalore, India 560 080}\\*[1cm]
\end{center}

\noindent
{\small We present an interesting connection between Brownian motion 
and magnetism. We use this to determine the distribution of areas 
enclosed by the path of a particle diffusing on a sphere. In addition, 
we find a bound on the free energy of an arbitrary system of spinless 
bosons in a magnetic field. The work presented here is expected to shed 
light on polymer entanglement, depolarized light scattering, and 
magnetic behavior of spinless bosons. } \\~\\*[0.8cm]

\noindent
Consider a particle diffusing on a sphere. If the diffusing particle 
returns to its starting point at time $\beta$ its path subtends a solid angle 
$\Omega$ at the center of the sphere. We ask: {\it what is the probability 
distribution of} $\Omega$ ? This problem comes up if one considers a 
spin-$\frac{1}{2}$ system in a random magnetic field. As is well known, 
the state (up to a phase) of a spin-$\frac{1}{2}$ system can be represented as a 
point on the Poincar\`e sphere. Under the influence of a random 
Hamiltonian, the state of the system diffuses on the Poincar\`e sphere. 
From the work of Berry$^{1}$ and others, it is known that the system picks 
up a geometric phase $\gamma$ equal to half the solid angle swept out 
on the Poincar\`e sphere. To compute the distribution of geometric phases 
one is led to the question posed above. A closely related problem has 
already been studied in the context of polymer entanglement:$^{2}$ given 
that a Brownian path on the plane is closed at time $\beta$, what is the 
probability that it encloses a given area $A$? \\

\noindent
In this paper we present a general method of solving these problems by 
using a connection between Brownian motion and magnetism. The qualitative 
idea is to use a magnetic field as a ``counter," to measure the area 
enclosed in a Brownian motion. We derive a relation between the 
distribution of areas in a Brownian motion and the partition function 
of a magnetic system, which can be used to cast light on both subjects. 
Despite its apparent simplicity, this relation does not seem to have 
been noticed or exploited so far. Our main purpose here is to 
illustrate its usefulness. We first discuss the planar problem solved 
earlier. We then go on to solve the (as yet unsolved) problem of 
diffusion on the sphere. We also exploit the relation to learn about 
the magnetic properties of bosonic systems. Here we recover previously 
known results and arrive at some others. We conclude the paper with 
a few remarks. \\

\noindent 
Let a diffusing particle start from a point on a plane at time $\tau = O$. 
Given that the path is closed at time $\beta$ (not necessarily for the 
first time), what is the conditional probability that it encloses a 
given area $A$? By ``area" we mean the algebraic area, including sign. 
The area enclosed to the left of the diffusing particle counts as 
positive and the area to the right as negative. This problem has been 
posed and solved$^{2}$ by polymer physicists, since it provides an 
idealized model for the entanglement of polymers. We present a method 
of solving this simple problem. \\

\noindent 
Let $\{\vec{x}(\tau), 0 \leq \tau \leq \beta, \vec{x}(0)=\vec{x}(\beta)\}$ be any 
realization of a closed Brownian path on the plane. As is well known, 
Brownian paths are distributed according to the Wiener measure:$^{3}$ 
if $f[\vec{x}(\tau)]$ is any functional on paths, the expectation value of 
$f$ is given by 
\begin{eqnarray}
&&\langle f[\vec{x}(\tau )]\rangle_{\mathcal{W}}\nonumber\\
&&\hspace*{1cm}\equiv \frac{\int {\cal{D}}[\vec{x}(\tau)]f[\vec{x}(\tau)]\exp \left[ -\int^{\beta}_{0}\left\{ \frac{1}{2} \frac{d\vec{x}}{d\tau} \cdot \frac{d\vec{x}}{d\tau} d\tau \right\}\right]}
{\int{\cal{D}}[\vec{x}(\tau)]\exp\left[ -\int^{\beta}_{0}\left\{ \frac{1}{2} \frac{d\vec{x}}{d\tau} \cdot \frac{d\vec{x}}{d\tau} d\tau \right\}\right]}.
\end{eqnarray}

\noindent
In Eq. (1) the functional integrals$^{4}$ are over all closed paths 
(the starting point is also integrated over). (We set the diffusion 
constant equal to half throughout this paper.) Let ${\cal{A}}[\vec{x}(\tau)]$
be the algebraic area enclosed by the path $\vec{x}(\tau)$. Clearly, the 
normalized probability distribution of areas {\it P(A)} is given by 
\begin{equation}
P(A) \equiv \langle\delta ({\cal {A}}[\vec{x}(\tau)]-A)\rangle_{\cal {W}}.
\end{equation}

\noindent
The expectation value $\tilde{\phi}$ of any function $\phi(A)$ of the area is 
given by $\int P(A)\phi(A)d(A)$. As is usual in probability theory we focus 
on the generating function $\tilde{P}(B)$ of the distribution $P(A)$:
\begin{equation}
\tilde{P}(B) \equiv {\overline{e^{ieBA}}} = \int P(A)e^{ieBA} dA,
\end{equation} 
which is simply the Fourier transform of $P(A)$. For future convenience we 
write the Fourier transform variable as $eB$. The distribution $P(A)$ can be 
recovered from its generating function by an inverse Fourier transform. 
From Eqs. (2) and (3) above we find 

\begin{equation}
\tilde{P}(B) = \langle e^{ieB{\cal{A}}}\rangle_{{\cal{W}}}.
\end{equation}
Notice that $B{\cal{A}}$ can be expressed as 
\begin{equation}
B{\cal{A}} = \int^{\beta}_{0} \vec{A}(\vec{x}) \cdot \frac{d\vec{x}}{d\tau} d\tau ,
\end{equation}
where $\vec{A}(\vec{x})$ is any vector potential whose curl is a homogeneous 
magnetic field $B$. Equations (1), (4), and (5) yield 
\begin{equation}
\tilde{P}(B) = \frac{\int{\cal{D}}[\vec{x}(\tau)]\exp\left[ \int^{\beta}_{0}\left\{
- \frac{1}{2} \frac{d\vec{x}}{d\tau}\cdot \frac{d\vec{x}}{d\tau} d\tau\right\} +
ie \int^{\beta}_{0} \left\{ \vec{A} \cdot \frac{d\vec{x}}{d\tau} d\tau \right\}\right]}
{\int {\cal{D}}[\vec{x}(\tau)]\exp\left[ \int^{\beta}_{0} \left\{ -\frac{1}{2} \frac{d\vec{x}}{d\tau} \cdot \frac{d\vec{x}}{d\tau} d\tau \right\}\right]}.
\end{equation}
By inspection of Eq.(6) we arrive at
\begin{equation}
\tilde{P}(B) = Z(B)/Z(0), 
\end{equation} 
where $Z(B)$ is the partition function $(Z(B) = Tr\{\exp[-\beta H(B)]\})$ 
for a quantum particle of charge e in a homogeneous magnetic field $B$ at an 
inverse temperature $\beta$. This is the central result of this paper and it 
relates Brownian motion and magnetism. As the reader can easily verify, the relation 
(7) holds even if there is an arbitrary biasing potential. The plane can also be 
replaced by a sphere or $(\re^{3})^{N}$, the configuration space of $N$ particles 
in $(\re^{3})$. In the last case, the area of interest is the sum of the weighted 
areas of the projections of the closed Brownian paths onto the $x-$y plane. 
Now we demonstrate the utility of Eq. (7) by computing the distribution of 
areas for diffusion on a plane. The partition function $Z(B)$ for a particle 
of unit mass in a constant magnetic field, is easily computed from the 
energies $E_{n} = (n + \frac{1}{2})eB$ and degeneracy (or the number of states 
per unit area) $(eB/2\pi)$ of Landau levels$^{5}$ (throughout this paper we set 
$\hbar = c = 1)$: 
\begin{equation*}
Z(B) = \sum^{\infty}_{n=0} \frac{eB}{2\pi} \exp[ -(n+\frac{1}{2})\beta eB] = \frac{eB/4\pi}{\sinh (\beta eB/2)}.
\end{equation*}
From (7) we find $\tilde{P}(B) = (\beta eB/2)[\sinh (\beta eB/2)]^{-1}$. 
Taking the Fourier transform of $\tilde{P}(B)$ by contour integration we get the 
result ${P}(A) = (\pi/2\beta)[\cosh(\pi A/\beta)]^{-2}$ derived in Ref. 2. 
This provides a check on Eq. (7) and illustrates its use.\\

\noindent 
Let us now address the problem posed at the beginning of this paper: what is 
the distribution $P(\Omega)$ of solid angles enclosed by a diffusing particle 
on a unit sphere? Unlike the planar case, $P(\Omega)$ is a periodic$^{6}$ function 
with period $4\pi$. The generating function $P_{g}$ of the distribution of solid 
angles is given by 
\begin{equation}
\tilde{P}_{g} = \int^{4\pi}_{0} d\Omega P(\Omega)e^{i \frac{g\Omega}{2}}
\end{equation}
with $g$ an integer. $P(\Omega)$ is expressed in terms of $\tilde{P}_{g}$ 
by a Fourier series 
\begin{equation}
P(\Omega) = \frac{1}{4\pi} \sum^{\infty}_{g=-\infty} e^{-i\frac{g\Omega}{2}}\tilde{P}_{g}
\end{equation}
rather than an integral (3). Relation (7) now takes the form 
\begin{equation}
\tilde{P}_{g} = Z_{g}/Z_{0},
\end{equation}
where $Z_{g}$ is the partition function for a particle of charge $e$ 
on a sphere subject to a magnetic field created by a monopole of quantized 
strength $G = g/e$ (Ref. 7) at the center of the sphere. The energy levels of this 
system are easily computed:$^{8}$ 
\begin{equation*}
E_{j} = [j(+1) - g^{2}]/2,
\end{equation*}
where $j$, the total angular momentum quantum number ranges from $|g|$ to infinity, 
and the $j{\rm th}$ level is $(2j + 1)$-fold degenerate. The partition function is 
consequently given by
\begin{equation}
Z_{g} = \sum^{\infty}_{j=|g|} (2j + 1)e^{-\frac{\beta\{j(j+1)-g^{2}\}}{2}}.
\end{equation}
Combining (9),(10), and (11) and rearranging the summations we arrive at 
\begin{eqnarray}
&&P(\Omega) = {\rm Re} \frac{1}{2\pi Z_{0}} \sum^{\infty}_{l=0} \left\{(2l+1) \frac{1+\zeta}{2(1-\zeta)}\right.\nonumber\\
&&\hspace*{1.5cm}\left.+\left[ \frac{2\zeta}{(1-\zeta)^{2}} \right]\right\} e^{\frac{-\beta}{2} l(l+1)},
\end{eqnarray}
where $\zeta (l,\beta,\Omega)=\exp[-1/2\{\beta(2l+1)+i\Omega\}].$ 
The function (12) is plotted numerically for various values of $\beta$ in Fig. 1. 
The qualitative nature of these plots is easily understood. For small values of 
$\beta$ the particle tends to make small excursions and its path encloses 
solid angles close to 0 or $4\pi$ and consequently the plots are peaked around 
these two values. As the available time $\beta$ increases, other values of 
$\Omega$ are also probable and the peaks tend to spread and the curves to flatten 
out. Finally in the limit of $\beta\rightarrow\infty$ the particle has enough 
time to enclose all possible solid angles with equal probability. These plots 
give the answer to the question that was raised in the beginning of the paper.\\

\noindent
Now we turn to the magnetic properties of spinless bosons. An $N$ particle 
system in three dimensions placed in a homogeneous external magnetic field 
which is along the $z$ direction has the Hamiltonian \\
\begin{center}
\includegraphics*[width=3.5in]{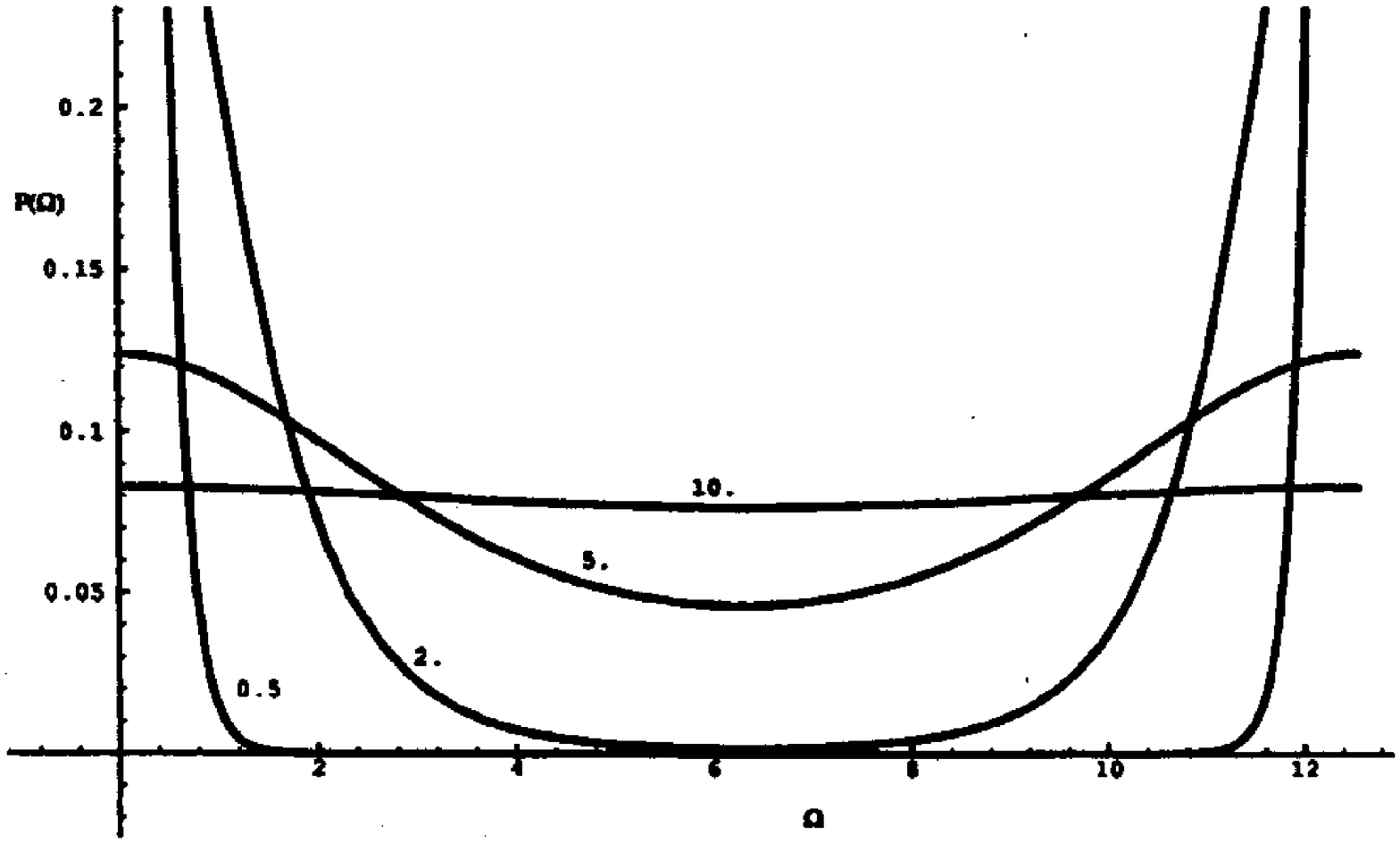}
\end{center}
\vspace*{0.3cm}

\begin{center}
{\small\sf  FIG.1.~~The probability distribution $P(\Omega)$ of solid angles for closed
randon walks lasting a time $\beta$. $P(\Omega)$ is plotted above for
four values of $\beta$:0.5, 2, 5, and 10.}
\end{center}

\begin{equation}
H(\vec{x^{a}}, \vec{p_{a}}) = \sum^{N}_{a=1} \frac{[\vec{p_{a}}-e_{a}\vec{A}(\vec{x^{a}})]^{2}}{2m_{a}} + V(\vec{x^{a}}),\nonumber
\end{equation}
where $\vec{A}(\vec{x^{a}})$ is the vector potential of the external magnetic 
field. $V(\vec{x_{a})}$ includes an arbitrary interaction between the particles 
as well as an external potential, $m_{a}$ and $e_{a}$ are the masses and charges 
of the particles. \( \{ \vec{x^{a}}, a = 1,2, \ldots ,N\} \) are the position vectors 
of the $N$ particles. The configuration space of the system is given by 
$Q=({\re}^{3})^{N}/\sim$, where $\sim$ means that we identify points in 
$({\re}^{3})^{N}$ which differ by an exchange of identical particles. 
For simplicity we give the argument for $N$ identical particles with unit mass 
and charge $e_{a}=e$. The argument is easily adapted to several species of particles 
of arbitrary charge and mass.\\

\noindent 
Now consider a diffusion on $Q$ biased by the potential $V(\vec{x^{a}})$. 
The Wiener measure is now appropriately modified:
\begin{equation*}
\langle f[\vec{x}(\tau)]\rangle_{{\cal{W}}(V)} \equiv
\frac{\int {\cal{D}}[\vec{x}(\tau)]f[\vec{x}(\tau)]\exp\left[-\int^{\beta}_{0}\left\{\frac{1}{2}\left( \sum_{a}\frac{d\vec{x_{a}}}{d\tau} \cdot \frac{d\vec{x_{a}}}{d\tau}\right)+V(\vec{x_{a}})\right\}d\tau\right]}
{\int {\cal{D}}[\vec{x}(\tau)]\exp\left[-\int^{\beta}_{0}\left\{\frac{1}{2}\left( \sum_{a} \frac{d\vec{x_{a}}}{d\tau} \cdot \frac{d\vec{x_{a}}}{d\tau}\right) + V(\vec{x^{a}})\right\} d\tau \right]}.
\end{equation*}

\noindent
The area whose distribution we are interested in is defined as follows: Let $q(\tau)$ 
be a closed curve in $Q\cdot q(\tau)$ determines trajectories of particles 
\(\{\vec{x^{a}}(\tau), a=1,2,\ldots,N\} \) in ${\re}^{3}$. The area functional 
of interest is 
${\cal{A}}[q(\tau)]\equiv \sum_{a}\int\vec{A}(\vec{x^{a}})\cdot d\vec{x^{a}}$. 
The area functional has the following interpretation. If the final positions of the 
$N$ particles are the same as the initial ones (direct processes), 
${\cal{A}}[q(\tau)]$
is simply the sum of the areas enclosed by the projection of the particle 
trajectories on the $(x-y)$ plane. If the final positions differ from the initial 
ones by a permutation (exchange processes), the projections of the particle 
trajectories still define closed curves on the $(x-y)$ plane. ${\cal{A}}[q(\tau)]$ 
is defined as the sum of areas enclosed by these closed curves.\\

\noindent 
As before we find that $\tilde{P}(B)$, which is the Fourier transform of the 
distribution $P(A) \equiv \langle\delta ({\cal{A}}[\vec{x}(\tau)]-A)\rangle_{{\cal{W}}(V)}$
of areas, is given by Eq. (7). It is crucial for our argument that the particles 
obey Bose statistics.$^{9}$ Since $P(A)$ is a probability distribution, 
$\tilde{P}(B) = Z(B)/Z(O)$ is the Fourier transform of a positive function. 
This places strong restrictions on the partition function $Z(B)$. Let 
$u_{i}, i=1,\ldots,n$ be $n$ real numbers. If one defines the $n \times n$ 
matrix $D^{(n)}_{ij} = \tilde{P}(u_{i}-u_{j})$, the necessary and sufficient 
condition for $\tilde{P}(B)$ to be the Fourier transform of a positive 
function is$^{10,11}$
\begin{equation}
\Delta^{(n)} \equiv {\rm Det} D^{(n)} \geq 0\;\;\;{\rm for\;\; all}\;\;n.
\end{equation}
This imposes restrictions on the free energy $F(B) = -(1/\beta)\ln Z(B)$ of the 
system in the presence of a magnetic field $B$. \\

\noindent
For the simplest nontrivial case $n = 2$, the inequality (13) with 
$B = u_{1} - u_{2}$ leads to 
\begin{equation}
Z(B) \leq Z(0)
\end{equation}
or equivalently, $F(B) \geq F(0)$. Since the free energy of the system increases 
in the presence of a magnetic field, the material is diamagnetic. This universal 
diamagnetic behavior of spinless bosons at all temperatures is known in the 
mathematical physics literature.$^{11}$ However, our approach may be accessible to 
a wider community of physicists. Our approach relating Brownian motion to magnetism 
enriches both fields and provides each field with intuition derived from the 
other. For instance, the zero-field susceptibility 
$\chi = -\partial^{2}F(B)/\partial B^{2}|_{B=0}$ of the magnetic system is related 
to the variance of the distribution of areas in the diffusion problem: 
\begin{equation}
\chi = \frac{1}{\beta}[\ln\tilde{P}(B)]^{\prime\prime}|_{B=0} = - \frac{e^{2}}{\beta}\overline{(A-\bar{A})^{2}} = -\frac{e^{2}}{\beta}{\rm Var} A\leq 0.
\end{equation}
It is curious that the zero-field susceptibility can be interpreted as the 
variance of the distribution of areas. Since the variance cannot be negative, 
it follows that $\chi$, the zero-field susceptibility cannot be positive and so 
these systems are diamagnetic.\\

\noindent 
Next consider the case $n = 3$. The $3 \times 3$ matrix $D^{(3)}_{ij}$ will then 
be a function of $u = u_{1} - u_{2}$ and $v = u_{2} - u_{3} (u_{1} - u_{3}$, 
being expressible in terms of $u$ and $v)$. If we set $u = 0$ 
(i.e., set $u_{1} = u_{2} =0, u_{3} = -v)$, we find that 
$\Delta^{(3)}(u,v)|_{u=0} = 0$. It then follows from the inequality (13) that 
$\Delta^{(3)}(u,v)$ has a minimum at $u = 0$ for all $v$. This implies that 
$\partial^{2}\Delta^{(3)}/\partial u^{2}|_{u=0, v=B} \geq 0$. Defining the 
function ${\cal{U}}(B) = \tilde{P}^{\prime 2}[1-\tilde{P}^{2}]^{-1}$, where the 
prime means derivative with respect to the magnetic field, we find
\begin{equation}
{\cal{U}}(B) \leq {\cal{U}}(0),
\end{equation} 
As can be seen by taking the limit $B \rightarrow 0$, ${\cal{U}}(0) = -\tilde{P^{\prime\prime}}(0) = - \beta\chi(0)$. We define a critical field 
$B_{c} = \pi/[2\sqrt{-\beta\chi(0)}]$. The inequality (16) implies a bound on the 
partition function. Notice that $\tilde{P}^{\prime}$ lies in a cone defined by the 
lines of slope $-(\pi/2B_{c})\sqrt{(1-\tilde{P}^{2})}$ and $(\pi/2B_{c})\sqrt{(1-\tilde{P}^{2})}$. It follows that
\begin{equation}
\tilde{P}(B) \geq \cos(\pi B/2B_{c})\;\;{\rm for}\;\;|B|\leq B_{c}.
\end{equation} 
The diamagnetic inequality due to Simon and Nelson$^{11}$ gives an upper bound 
on the partition function $Z(B)$ of a system of spinless bosons. The new 
inequality stated in (17) gives us a lower bound on $Z(B)$ (see Fig. 2) (or 
equivalently, an upper bound on the free energy).\\
\begin{center}
\includegraphics*[width=4.5in]{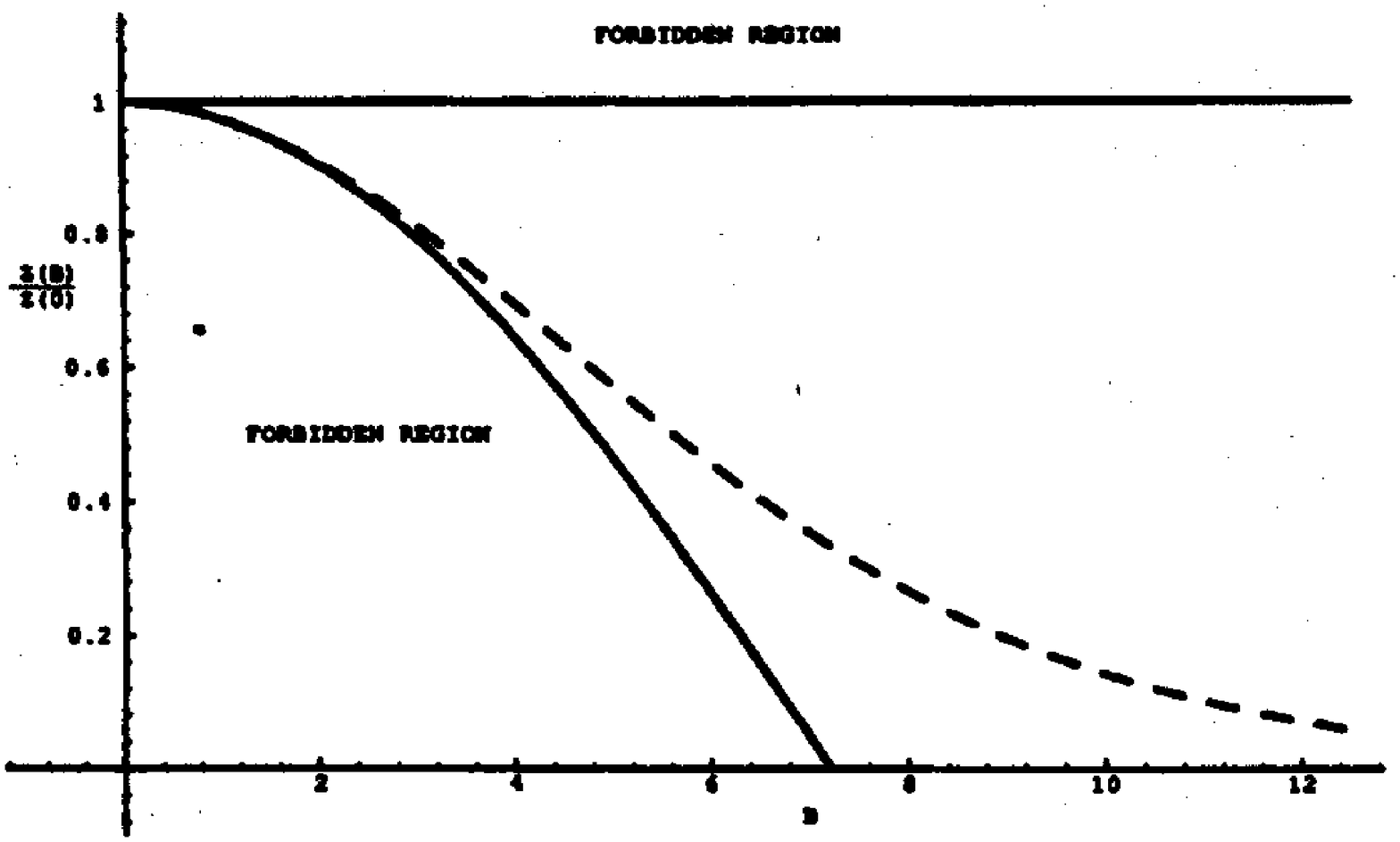}
\end{center}
\vspace*{0.3cm}

\begin{center}
{\small\sf Fig. 2. The region forbidden by the bounds [inequalities (14) and (18) on the 
partition function. These bounds are shown as solid lines. The dotted curve is 
the partitition for a charged simple harmonic oscillator in an external magnetic 
field. Notice that the dotted curve lies outside the forbidden region.}
\end{center}
\vspace*{0.4cm}

\noindent 
As an explicit check on this new bound on the free energy we considered a simple 
system--a charged particle in a magnetic field subject to a harmonic oscillator 
potential. The calculated partition function of this system is close to, but 
above the lower bound set by (17). Needless to say, our bound is derived for 
an arbitrary interacting system of spinless bosons. The new bound presented 
here along with the earlier (14) diamagnetic inequality$^{11}$ places strong 
restrictions on the partition function of a bosonic system in the presence of 
a magnetic field. We find a curious and immediate consequence of these 
restrictions: if the zero-field susceptibility of the system vanishes, then 
Eqs. (14) and (17) imply that $Z(B) = Z(O)$, i.e., the system is nonmagnetic 
at all fields. \\

\noindent
The key result of this paper is a connection between two apparently distinct 
classes of problems--Brownian motion and magnetism. This allows us to compute 
the distribution of solid angles enclosed in Brownian motion on a sphere. 
As mentioned earlier, this problem comes up when computing the distribution of 
Berry phases in a random magnetic field. A more classical context is depolarized 
light scattering. As is well known, a light ray following a space curve picks 
up a geometric phase,$^{12,13}$ equal to the solid angle swept out by the 
direction vector. If a light ray inelastically scatters off a random medium, 
its direction vector does a random walk on the unit sphere of directions. The 
distribution $P(\Omega)$ computed here is relevant to the extent of 
depolarization in such an experiment.$^{14}$\\

\noindent
In the domain of magnetism we find an independent way of arriving at the 
diamagnetic inequality$^{11}$ which states that the free energy of a system of 
spinless bosons always increases in the presence of a magnetic field. Spinless 
charged bosonic systems occur in the context of superconductors (which are 
perfect diamagnets) and neutron stars.$^{15}$ We believe that the community 
of physicists working in these areas may not be aware of the general results 
available in the mathematical literature. For instance, the diamagnetism of 
bosons may be relevant$^{16}$ to the interpretation of recent experiments$^{17}$ 
on high-$T_{c}$ superconductivity. \\

\noindent
Throughout this paper we have only discussed homogeneous magnetic fields. It is 
easy to generalize our discussion to take into account arbitrary inhomogeneous 
fields: all one does is consider the distribution of {\it weighted} areas. An 
obvious application of this is the computation of the probability of 
entanglement of a polymer with a background lattice of polymers. We expect 
the new method outlined here to shed light on open problems in polymer 
entanglement involving more complicated configurations of polymers than the 
simplest one solved so far. One can also use the relation (7) to compute the 
distribution of winding numbers in diffusion in a multiply connected space.\\

\noindent 
It is a pleasure to thank N. Kumar for bringing up the problem of diffusion 
on a sphere and several discussions on this work; Barry Simon for his help 
in finding Ref. 11; Diptiman Sen for discussions and for giving us Ref. 8, 
and R. Nityananda for discussions and drawing our attention to Ref. 10.\\~\\

\begin{center}
$------------$
\end{center}
\newpage

\begin{itemize}
\item[$^{1}$] See Geometric Phases in Physics, edited by Alfred Shapere and 
              Frank Wilczek (World Scientific, Singapore, 1989).

\item[$^{2}$] M. G. Brereton and C. Butler, J. Phys. (London) A {\bf 20}, 3955, 
              (1987); 
              D. C. Khandekar and F. W. Wiegel, J. Phys. (London) A {\bf 21}, 
              L563 (1988); 
              J. Stat. Phys. {\bf 53}, 1073 (1988). 

\item[$^{3}$] L. S. Schulman, Techniques and Applications of Path Integration, 
              (Wiley, New York, 1981). 

\item[$^{4}$] R. P. Feynman and A. R. Hibbs, Quantum Mechanics and Path Integrals 
              (McGraw-Hill, New York, 1965). 

\item[$^{5}$] L. D. Landau and E. M. Lifshitz, Quantum Mechanics (Pergamon, Oxford, 
              1977), p. 457. 

\item[$^{6}$] A path that encloses a solid angle $\Omega$ to its left also 
              encloses a solid angle of $4\pi-\Omega$ to its right. Using the 
              symmetry $P(\Omega) = P(-\Omega)$ we conclude that $P(\Omega)$ 
              is periodic with period $4\pi$.

\item[$^{7}$] P. A. M. Dirac, Proc. R. Soc. London {\bf A133}, 60 (1931); 
              M. N. Saha, Ind. J. Phys. {\bf 10}, 141 (1936). 

\item[$^{8}$] S. Coleman, in {\it The Unity of Fundamental Interactions}, 
              edited by A. Zichichi (Plenum, New York, 1983), p. 21-117. 

\item[$^{9}$] The identity (10) is not true for Fermi particles since in this 
              case exchange processes are weighted negatively relative to direct 
              ones in the expression for the partition function. This ruins the 
              probability interpretation that we give to the partition function. 

\item[$^{10}$]M. Komesaroff et al., Astron. Astrophys. {\bf 93}, 269 (1981).
 
\item[$^{11}$]B. Simon, Phys. Rev. Lett. {\bf 36}, 1083 (1976); 
              Indiana Univ. Math. J. {\bf 26}, 1067 (1977); 
              J. Functional Anal. {\bf 32}, 97 (1979); 
              {\it Functional Integration and Quantum Physics} (Academic, 
              New York, 1979), and references therein. 

\item[$^{12}$]L. D. Landau and E. M. Lifshitz, Electrodynamics of Continuous 
              Media (Pergamon, Oxford, 1960), p. 271. 

\item[$^{13}$]A. Tomita and R. Chiao, Phy. Rev. Lett. {\bf 57}, 937 (1986). 

\item[$^{14}$]S. Sanyal et al., Phys. Rev. Lett. {\bf 72}, 2963 (1994). 

\item[$^{15}$]J. Daicic et al., Phys. Rep. {\bf 237}, 65 (1994); 
              J. Daicic et al., Phys. Rev. Lett. {\bf 71}, 1779 (1993). 

\item[$^{16}$]In these experiments, diamagnetism is observed well above $T_{c}$.
              A possible explanation may be that the diamagnetism is due to 
              the persistence of Cooper pairs (spinless bosons) well above 
              $T_{c}$. We thank N. Kumar for this observation. See also 
              A. S. Alexandrov and N. F. Mott, Phys. Rev. Lett. {\bf 71}, 
              1075 (1993); D. C. Bardos et al., Phys. Rev. B {\bf 49}, 4082 (1994). 

\item[$^{17}$]W. C. Lee et al., Phy. Rev. Lett. {\bf 63}, 1012 (1989). 
\end{itemize}
\end{document}